\documentclass[journal=jctcce,manuscript=article]{achemso}
\setkeys{acs}{usetitle=true}

\usepackage[version=4]{mhchem}
\usepackage{graphicx, subcaption}
\usepackage{booktabs,multirow,multicol,tabularx,threeparttable}
\usepackage[table,xcdraw]{xcolor}
\usepackage{amsmath,amssymb,bm,mathtools,braket}
\usepackage[ruled,vlined,linesnumbered]{algorithm2e}

\usepackage{siunitx}
\usepackage{hyperref}
\usepackage{cleveref}
\usepackage{float}

\usepackage{listings}
\title{A High-Performance Pauli-Algebra Framework for Large-Scale Quantum Simulations}

\author{Xiaopeng Li}
\affiliation{State Key Laboratory of Precision and Intelligent Chemistry, University of Science and Technology of China, Hefei, Anhui 230026, China}
\author{Zhijie Sun}
\affiliation{State Key Laboratory of Precision and Intelligent Chemistry, University of Science and Technology of China, Hefei, Anhui 230026, China}
\author{Yi Fan}
\affiliation{Hefei National Research Center for Physical Sciences at the Microscale, University of Science and Technology of China, Hefei, Anhui 230026, China}
\author{Jie Liu}
\email{liujie86@ustc.edu.cn}
\affiliation{Hefei National Laboratory, University of Science and Technology of China, Hefei 230088, China}
\author{Zhenyu Li}
\email{zyli@ustc.edu.cn}
\affiliation{Hefei National Laboratory, University of Science and Technology of China, Hefei 230088, China}
\alsoaffiliation{State Key Laboratory of Precision and Intelligent Chemistry, University of Science and Technology of China, Hefei, Anhui 230026, China}
\author{Jinlong Yang}
\affiliation{Hefei National Laboratory, University of Science and Technology of China, Hefei 230088, China}
\alsoaffiliation{State Key Laboratory of Precision and Intelligent Chemistry, University of Science and Technology of China, Hefei, Anhui 230026, China}


\begin{document}

\begin{abstract}
We present a high-performance Pauli-algebra framework for large-scale quantum chemistry and quantum many-body simulations. In this framework, products and linear combinations of Pauli strings evaluated directly in a compact binary symplectic representation using hardware-efficient bitwise operations and canonical reduction, thereby avoiding costly symbolic manipulation of Pauli strings. More importantly, operator--state multiplication is accelerated by Pauli-$X$-mask grouping combined with target-configuration-based summation and symmetry blocking. This scheme reduces configuration transition loops, improves data locality and memory efficiency, and enables efficient parallel operator applications. Numerical benchmarks demonstrate Jordan–Wigner transformations with more than two orders of magnitude speedup, large-active-space VQE calculations of molecular potential energy surfaces up to 56 qubits, and efficient real-time quantum dynamics on modern multicore CPU and GPU architectures. These results demonstrate that exploiting the algebraic and sparsity structures of Pauli representations provides a practical route toward scalable classical emulation and benchmarking of quantum algorithms for quantum chemistry and many-body simulations.

\end{abstract}

\section{Introduction}
The rapid development of quantum hardware has greatly accelerated the search for quantum algorithms capable of addressing challenging problems in chemistry and materials science.\cite{cao2019quantum,McAEndAsp20,bauer2020quantum,LiuFanLi22} Among these efforts, the variational quantum algorithm, has emerged as a leading near-term approach for electronic structure calculations\cite{peruzzo2014variational,Cerezo2021,Tilly_VQE_2021,LiuWanLi20} and dynamics simulations\cite{yuan2019theory,Miessen2023QuantumDynamics,Linteau2024AdaptivePVQD}, while quantum phase estimation (QPE)\cite{abrams1999quantum,kitaev1995quantum,aspuru2005simulated} provide a systematic route toward high-accuracy energy estimation on future fault-tolerant devices. At the same time, classical simulation remains indispensable for designing, testing, and benchmarking quantum algorithms.\cite{pyquil,smelyanskiy2016qhipster,projectq,bergholm2018pennylane,qsim,jones2019quest,qulacs,mcclean2020openfermion,JUSTC-2022-0118,WilWilJin22,cuquantum,qiskit-package,LiAllChe23} However, faithfully emulating quantum algorithms on classical computers is highly challenging because the Hilbert-space dimension grows exponentially and because algorithmic primitives such as variational ansatz preparation, controlled time evolution, and dynamical propagation require repeated manipulation of high-dimension quantum states.

Recently, substantial effort has been directed toward the classical emulation of quantum algorithms as an essential step before their reliable implementation on large-scale quantum hardware. One direct strategy is state-vector simulation, where the complete quantum state is explicitly represented and evolved under quantum gates. Although this approach is both general and exact, it suffers from memory and computational requirements that grow exponentially with the number of qubits.\cite{smelyanskiy2016qhipster,Luo2020Yao,jones2019quest,qulacs,isakov2021qsim} Specialized fermionic state-vector simulators, e.g. FQE~\cite{FQE} and ffsim~\cite{Sung2026ffsim} that work in fermionic Fock spaces, have been proposed to reduce memory and computational costs by exploiting particle-number and spin symmetries. Tensor-network methods provide an alternative route by exploiting circuit structure, locality, and limited entanglement, enabling efficient simulations for classes of low-entanglement states or circuits with favorable contraction structures.\cite{Vidal2003,Markov2008,Gray2021,Pan2022} These approaches have become powerful tools for analyzing quantum circuits and many-body dynamics, but their efficiency can depend strongly on entanglement growth, circuit connectivity, and contraction-path optimization. 

More recently, algebraic simulation approaches have also been developed to exploit the restricted operator structure of certain quantum circuits. For example, Lie-algebraic simulation methods, such as $\mathfrak{g}$-sim, avoid explicit state-vector propagation by tracking the evolution of observables within a dynamical Lie algebra, which can be efficient when the relevant algebra has polynomial dimension.\cite{goh2025lie,BarSimKot26} These methods highlight the value of exploiting algebraic structure in classical emulation, but their efficiency depends on the closure and dimension of the underlying operator algebra.\cite{kokcu2022fixed,wan2024hybrid,wan2025calculation} In addition, stabilizer-based simulators provide another efficient route for Clifford-dominated circuits.\cite{gottesman1997stabilizer,gidney2021stim} In quantum chemistry algorithms, where molecular Hamiltonians and chemically inspired ans\"atze are naturally expressed in terms of large collections of Pauli strings and their exponentials, a Pauli-algebraic simulation framework provides a complementary route for accelerating repeated state transformations. Here, Pauli-algebraic objects refer to computational objects represented or manipulated within the algebra generated by Pauli strings, including Pauli-sum Hamiltonians and observables, excitation generators, operator products and commutators, gradients, and exponentiated unitary factors arising in variational ansatz preparation and time evolution.

The binary symplectic representation provides a compact and efficient encoding of Pauli strings and has been extensively used in stabilizer theory, Clifford-circuit simulation, qubit tapering, and Pauli-operator manipulation.\cite{gottesman1997stabilizer,dehaene2003clifford,bravyi2017tapering,gidney2021stim,pauliarray} In this representation, each $n$-qubit Pauli string is specified by two binary vectors corresponding to its Pauli-$X$- and Pauli-$Z$-type components, with the overall phase stored separately. Consequently, Pauli multiplication, commutation checking, and phase updates can be reduced to bitwise operations and binary inner products, avoiding explicit symbolic manipulation of Pauli strings. Previous studies have exploited this representation mainly for stabilizer-tableau updates, Clifford simulation, symmetry reduction, and compact storage and manipulation of Pauli-operator arrays.\cite{gottesman1997stabilizer,dehaene2003clifford,gidney2021stim,bravyi2017tapering,pauliarray,hantzko2024tensorized}Despite these advances, most existing simulation frameworks are designed primarily around general quantum circuits, dense state-vector evolution, or restricted circuit classes. 

In quantum chemistry and many-body simulations, however, the central computational objects are often large Pauli-operator collections, including molecular Hamiltonians that may contain tens of thousands to millions of Pauli terms. In this setting, the computational bottleneck is not only the evolution of a dense state, but also the repeated algebraic manipulation of Pauli operators, Hamiltonian--state multiplications, and the exploitation of sparsity and symmetry-restricted configuration spaces.\cite{verteletskyi2020measurement} General-purpose circuit simulators typically do not fully exploit the algebraic structure, coefficient organization, and grouped sparsity inherent in such Hamiltonians, leading to unnecessary memory traffic and reduced computational efficiency. Recent efforts on efficient Pauli-string manipulation further highlight the importance of specialized operator-level engines for Pauli-structured quantum algorithms.\cite{pauliarray,hantzko2024tensorized}

To address these challenges, we present a high-performance Pauli-algebra simulation framework tailored to large-scale quantum chemistry and many-body quantum simulations. The efficiency of this framework originates from exploiting Pauli-algebraic structure at two complementary levels. First, Pauli multiplications and linear combinations are carried out directly in a compact binary symplectic representation, where phase tracking, operator products, and duplicate-term accumulation are reduced to hardware-efficient bitwise operations and Pauli-$(X,Z)$-mask-based addressing. This avoids costly symbolic Pauli manipulation and provides an efficient route for fermion-to-qubit mappings such as the Jordan--Wigner transformation. Second, for operator--state multiplication, Pauli strings are organized according to shared Pauli-$X$ masks, which separates configuration transitions from parity-phase evaluations. Together with a spin-resolved product basis and a target-configuration-based summation, it improves data locality, reduces redundant memory routing, and enables efficient parallel Hamiltonian-state multiplication in symmetry-adapted many-electron spaces. The resulting Julia/C\texttt{++} implementation, further accelerated on multicore CPUs and GPUs, provides a scalable Pauli-algebra backend for Hamiltonian construction, VQE calculations, and real-time quantum dynamics simulations.

\section{Methods}
\label{sec:methods}

\subsection{Quantum Chemistry and Many-Body Simulations}

Solving the Schr\"odinger equation lies at the heart of quantum chemistry and many-body simulation, and its implementation on digital quantum computers requires mapping the target Hamiltonian onto the qubit representation that can be realized and manipulated by quantum circuits. A general Hamiltonian in the qubit representation can be written as a Pauli-structured operator,
\begin{equation}
\hat{H} =
\sum_{j=1}^{M}
c_j \hat{P}_j ,
\label{eq:pauli_hamiltonian}
\end{equation}
where $\hat{P} \in \{I,\sigma_x,\sigma_y,\sigma_z\}^{\otimes n}$ denotes an $n$-qubit Pauli string. $I$ is the identity operator, and $\sigma_x$, $\sigma_y$, and $\sigma_z$ denotes Pauli-$X$, Pauli-$Y$ and Pauli-$Z$ operators, respectively. This representation provides a common algebraic language for a broad class of quantum simulation problems.

For quantum chemistry simulations, one usually starts from the second-quantized Hamiltonian,
\begin{equation}
\hat{H}_{\mathrm{elec}}
=
\sum_{pq} h_{pq} \hat{a}_p^\dagger \hat{a}_q
+
\frac{1}{2}
\sum_{pqrs}
h_{pqrs}
\hat{a}_p^\dagger \hat{a}_q^\dagger \hat{a}_r \hat{a}_s + E_{\mathrm{NN}},
\label{eq:second_quantized_hamiltonian}
\end{equation}
where $\hat{a}_p^\dagger$ and $\hat{a}_p$ are fermionic creation and annihilation operators, respectively. $h_{pq}$ and $h_{pqrs}$ are one- and two-electron integrals, and $E_{\mathrm{NN}}$ is the nuclear-nuclear repulsion energy. The fermionic operators can be mapped to Pauli operators through fermion-to-qubit transformations, such as the Jordan--Wigner,\cite{jordan1928pauli} parity,\cite{bravyi2002fermionic} or Bravyi--Kitaev\cite{seeley2012bravyi} mappings. After such a transformation, Eq.~\eqref{eq:second_quantized_hamiltonian} takes the form of Eq.~\eqref{eq:pauli_hamiltonian}. This Pauli-algebra representation is applicable to both finite molecular systems under open boundary conditions and periodic materials under periodic boundary conditions.\cite{LiuWanLi20,manrique2020momentum,yoshioka2022variational,LiuFanLi22,LiFanLiu25} 

On the other hand, many spin and lattice models are naturally expressed in terms of Pauli operators. For example, the transverse-field Ising model (TFIM) can be written as
\begin{align}\label{eq:TFIM}
\hat{H}_{\mathrm{TFIM}} =
J\sum_{\langle i,j\rangle} \sigma^z_{i} \sigma^z_{j}
+h\sum_{i} \sigma^x_{i},
\end{align}
where $\sigma^{x}_{i}$ ($\sigma^{y}_{i}$, $\sigma^{z}_{i}$) indicates that $\sigma_{x}$ ($\sigma_{y}$, $\sigma_{z}$) act on the $i$-th qubit. These models therefore already have the structure of Eq.~\eqref{eq:pauli_hamiltonian}, making Pauli-algebra operations a fundamental computational primitive not only for quantum chemistry but also for condensed-matter and model-Hamiltonian simulations.

In chemistry and materials science, quantum simulation tasks can be broadly divided into static and dynamical problems. Static quantum simulation focuses on the eigenvalue problem
\begin{equation}
\hat{H} |\psi_k\rangle = E_k |\psi_k\rangle ,
\label{eq:eigenvalue_problem}
\end{equation}
where the main objective is usually to obtain the ground- and/or excited-state energy $E_k$ and the corresponding eigenstate $|\Psi_k\rangle$. A leading algorithm for this purpose is the variational quantum eigensolver (VQE).\cite{peruzzo2014variational,Tilly_VQE_2021,Cerezo2021} In VQE, a parameterized quantum state $|\psi(\boldsymbol{\theta})\rangle$ is prepared by a quantum circuit, and the eigenenergy is estimated by minimizing the energy functional
\begin{equation}
E(\boldsymbol{\theta})
=
\langle \psi(\boldsymbol{\theta}) |
\hat{H}
|
\psi(\boldsymbol{\theta})
\rangle
=
\sum_{j=1}^{M}
c_j
\langle \psi(\boldsymbol{\theta}) |
\hat{P}_j
|
\psi(\boldsymbol{\theta})
\rangle .
\label{eq:vqe_energy}
\end{equation}

Dynamical quantum simulation, on the other hand, aims to simulate the time evolution of a quantum state under a given Hamiltonian. For real-time dynamics, the state evolves according to the time-dependent Schr\"odinger equation,
\begin{equation}
i\frac{\partial}{\partial t}
|\psi(t)\rangle
=
\hat{H}
|\psi(t)\rangle ,
\end{equation}
whose formal solution for a time-independent Hamiltonian is
\begin{equation}
|\psi(t)\rangle
=
e^{-i\hat{H}t}
|\psi(0)\rangle .
\label{eq:real_time_evolution}
\end{equation}
Since $\hat{H}$ is usually represented as a sum of noncommuting Pauli strings, implementing or classically emulating the propagator $e^{-i\hat{H}t}$ requires efficient manipulation of Pauli-algebra operators. Standard approaches include product-formula methods, such as Trotter--Suzuki decompositions,\cite{lloyd1996universal,suzuki1991general} as well as more advanced algorithms based on linear combination of unitaries, qubitization, quantum signal processing,\cite{berry2015simulating,low2019hamiltonian,low2017optimal} and cartan decomposition.\cite{Wan2024HybridHamiltonian,Wan2025GreenFunctionCartan}

Therefore, both static and dynamical quantum simulation ultimately rely on repeated algebraic operations involving a large batch of Pauli strings. In variational algorithms, these operations appear in Hamiltonian construction, expectation-value evaluation, gradient estimation, measurement grouping, and adaptive ansatz generation. In QPE and real-time simulation, they arise in the construction and application of time-evolution operators. Therefore, efficient manipulation of Pauli-algebraic objects is thus a central computational task in large-scale classical emulation and benchmarking of quantum algorithms. In this work, we focus on accelerating these Pauli-algebra operations by exploiting binary symplectic representations\cite{gottesman1997stabilizer,dehaene2003clifford,bravyi2017tapering,gidney2021stim,pauliarray}.

\subsection{Binary Symplectic Encoding}

\subsubsection{Single-Qubit Pauli Operator}

Each single-qubit Pauli operator can be mapped onto a pair of bits $(a, b)$ with $a,b \in \{0,1\}$, up to a global phase:
\begin{equation} \label{eq:single_pauli_mapping}
    \sigma^{(a,b)} = 
    \begin{cases}
        I, & (a,b) = (0,0), \\
        \sigma_x, & (a,b) = (1,0), \\
        \sigma_y, & (a,b) = (1,1), \\
        \sigma_z, & (a,b) = (0,1).
    \end{cases}
\end{equation}
In this convention, the pair $(1,1)$ represents $\sigma_y$. Since $\sigma_y=i\sigma_x\sigma_z$, the corresponding phase factor must be tracked separately when Pauli strings are multiplied. On a classical computer, the binary variables for multiple qubits can be packed into integers, which allows Pauli operations to be performed efficiently using bitwise instructions.

The Pauli type associated with a given bit pair $(a,b)$ can be recovered as
\begin{align}
    \mathcal{T}_y &= a \land b, \\
    \mathcal{T}_x &= a \oplus \mathcal{T}_y, \\
    \mathcal{T}_z &= b \oplus \mathcal{T}_y,
\end{align}
where $\oplus$ and $\land$ denote bitwise XOR and AND, respectively. Here, $\mathcal{T}_w=1$ indicates that the operator acting on the qubit is $\sigma_w$ for $w\in\{x,y,z\}$, while $\mathcal{T}_x=\mathcal{T}_y=\mathcal{T}_z=0$ corresponds to the identity operator.

\subsubsection{Multi-Qubit Pauli Strings}
An $n$-qubit Pauli string $\hat{P}$, including a global phase factor, can be written as\cite{dehaene2003clifford,gottesman1997stabilizer}
\begin{equation} \label{eq:pauli_group}
    \hat{P} = i^\gamma \bigotimes_{k=0}^{n-1} \sigma^{(a_k,b_k)},
\end{equation}
where the phase factor $\gamma \in \{0,1\}$. Instead of storing explicit $2^n \times 2^n$ matrices, one can employ binary symplectic encoding to represent this Pauli string with two bitstrings $X, Z \in \mathbb{N}$ and one coefficient $\Lambda=i^\chi$, denoted as $(X,Z,\Lambda)$. Consequently, the Pauli string can be recovered through 
\begin{equation}
    \hat{P}= \hat{\mathcal{R}}(X,Z,\Lambda) =
    \Lambda \bigotimes_{k=0}^{n-1}
    \left(\sigma_k^x\right)^{a_k}
    \left(\sigma_k^z\right)^{b_k},
\end{equation}
where
\begin{align}
    X &= \bigvee_{k=0}^{n-1} (a_k \ll k), \label{eq:xp_encoding} \\
    Z &= \bigvee_{k=0}^{n-1} (b_k \ll k), \label{eq:zp_encoding}
\end{align}
$\vee$ denotes bitwise OR, $\ll$ denotes bitwise left shift, and $\otimes$ denotes tensor product. The total phase, including both the explicit global phase $\gamma$ and the implicit phases from $\sigma_y$ operators, is $\chi = \gamma + \phi(X,Z)$, where $\phi$ counts the number of $\sigma_y$ operators, each contributing a factor of $i$, that is
    $\phi(X,Z) = \operatorname{popcnt}(X \land Z)$.
The function $\operatorname{popcnt}(\cdot)$ denotes the number of set bits, i.e., the number of bits equal to one, in the binary representation of an integer.

\subsection{Pauli-Algebra Operations: Linear Combination and Multiplication}
\label{sec:algebraic_ops}

A linear combination of $n$-qubit Pauli strings, such as the Hamiltonian $\hat{H}=\sum_{j=1}^M c_j \hat{P}_j$, can be represented by three arrays:
\begin{align}
    \mathbf{X} &= [X_1, X_2, \dots, X_M] , \\
    \mathbf{Z} &= [Z_1, Z_2, \dots, Z_M] , \\
    \mathbf{C} &= [C_1, C_2, \dots, C_M] ,
\end{align}
where $(X_j, Z_j, C_j)$ encodes $\hat{P}_j$ via \Cref{eq:xp_encoding,eq:zp_encoding}, and the coefficient $C_j = c_j*\Lambda_j$.

For two Pauli strings $\hat{P}_k$ and $\hat{P}_l$ encoded as $(X_k, Z_k, \Lambda_k)$ and $(X_l, Z_l, \Lambda_l)$, their product $\hat{P}_m = \hat{P}_k \hat{P}_l$ follows directly from the binary symplectic representation:
\begin{align}
    X_m &= X_k \oplus X_l, \label{eq:pauli_prod_x} \\
    Z_m &= Z_k \oplus Z_l, \label{eq:pauli_prod_z} \\
    \Lambda_m &= \Lambda_k \cdot \Lambda_l \cdot (-1)^{\phi(Z_k , X_l)}. \label{eq:pauli_prod_phase}
\end{align}
%The phase factor $(-1)^{\operatorname{popcnt}(Z_a \land X_b)}$ arises from anticommutation between $\sigma_z$ in $\hat{P}_a$ and $\sigma_x$ in $\hat{P}_b$ on overlapping qubits. 
The multiplication thus reduces to bitwise XOR, bitwise AND, population count, and scalar multiplication---all hardware-efficient operations on a classical computer.\cite{gidney2021stim}

For two Pauli-structured operators $\hat{A}=\sum_k a_k\hat{P}_k$ and $\hat{B}=\sum_l b_l\hat{P}_l$, their product is
\begin{equation}
    \hat{A}\hat{B}
    = \sum_{k,l} 
      \hat{\mathcal{R}}(X_k\oplus X_l,\; Z_k\oplus Z_l,\; C_{kl}),
    \label{eq:pauli_operator_multiplication}
\end{equation}
where $C_{kl}=a_k b_l\,\Lambda_k\Lambda_l \,(-1)^{\phi(Z_k, X_l)}$. Each pair $(k,l)$ produces one $(X,Z)$-mask by XOR and one coefficient. Different pairs may generate the same output mask, so the coefficients must be accumulated by binary key $(X_s\oplus X_t,\, Z_s\oplus Z_t)$.

Maintaining operators in canonical form---unique Pauli strings, sorted by $(X,Z)$, with accumulated coefficients---is essential for both efficiency and numerical stability. The canonical reduction procedure sorts all terms by their binary symplectic key using radix sort, yielding $\mathcal{O}(m\cdot n_q/w)$ complexity for $m$ terms on $n_q$ qubits with machine word width $w$. The sorted list is then scanned linearly to merge duplicate strings and accumulate their coefficients. A numerical threshold $\varepsilon_{\mathrm{drop}}$ can be applied to discard negligible terms.

\begin{algorithm}[!htb]
\SetKwData{DD}{dict}
\SetKwData{Key}{\ensuremath{k}}
\SetKwData{Val}{\ensuremath{C}}
\SetKwData{PP}{\ensuremath{p}}
\SetKwData{QQ}{\ensuremath{q}}
\SetKwData{RR}{\ensuremath{r}}
\SetKwData{SS}{\ensuremath{s}}
\SetKwData{TT}{\ensuremath{t}}
\SetKwFunction{Encode}{encode}
\SetKwFunction{Decodex}{decodeX}
\SetKwFunction{Decodez}{decodeZ}

\DontPrintSemicolon
\SetKwFunction{JWBuild}{JW\_Hamiltonian}
\SetKw{Function}{Function}
\Function{\JWBuild{$h_{pq}$, $h_{pqrs}$, $n$, $\varepsilon$}}\;
\KwIn{One- and two-electron integrals $h_{pq}$, $h_{pqrs}$; $n$ spin-orbitals; threshold $\varepsilon$}
\KwOut{Pauli Hamiltonian $\hat{H} = \sum_k c_k \hat{P}_k$ in canonical ordering}
\PrintSemicolon

\DD $\gets$ empty hash table keyed by $\Key = \Encode(X, Z)$\;
\For{\PP $\gets 0$ \KwTo $n-1$}{
    \For{\QQ $\gets 0$ \KwTo $n-1$}{
        \If{$|h_{{\PP}{\QQ}}| > \varepsilon$}{
            precompute $\left. (\Key_{\TT}, \Val_{\TT}) \right|_{\TT=1}^4$ for $(\hat{a}^\dagger_{\PP}\hat{a}_{\QQ})$ via Eq.~\eqref{eq:pauli_prod_x}--\eqref{eq:pauli_prod_phase}\;
            \For{\TT $\gets 1$ \KwTo 4}{
                $\DD[\Key_{\TT}] \gets \DD[\Key_{\TT}] + \Val_{\TT}$\;
            }
        }
        \For{\RR $\gets 0$ \KwTo $n-1$}{
            \For{\SS $\gets 0$ \KwTo $n-1$}{
                \If{$|h_{{\PP}{\QQ}{\RR}{\SS}}| > \varepsilon$}{
                    precompute $\left. (\Key_{\TT}, \Val_{\TT}) \right|_{\TT=1}^{16}$ for $(\hat{a}_{\PP}^\dagger\hat{a}_{\QQ}^\dagger\hat{a}_{\RR}\hat{a}_{\SS})$\;
                    \For{\TT $\gets 1$ \KwTo 16}{
                        $\DD[\Key_{\TT}] \gets \DD[\Key_{\TT}] + \Val_{\TT}$\;
                    }
                }
            }
        }
    }
}
sort entries of \DD by \bf{k}\;
output sorted list $\{(\mathbf{X}, \mathbf{Z}, \mathbf{C})\}$ as $\hat{H}$\;
\Return{$\hat{H}$}\;
\caption{Direct construction of the qubit Hamiltonian from electron integrals via Pauli multiplication and linear combination.}
\label{alg:jw_construction}
\end{algorithm}

Taking fermion-to-qubit transformations such as the Jordan--Wigner mapping as an example. Here, Pauli multiplication and canonical reduction are the computational primitives. The Jordan--Wigner transformation expresses each fermionic creation/annihilation operator as a product of Pauli operators:
\begin{equation}
    \hat{a}_p^\dagger = \tfrac{1}{2}(\sigma^x_p - i\sigma^y_p) \otimes \sigma^z_{p-1}\cdots \sigma^z_0,
    \qquad
    \hat{a}_p = \tfrac{1}{2}(\sigma^x_p + i\sigma^y_p) \otimes \sigma^z_{p-1}\cdots \sigma^z_0,
\end{equation}
where the phase string $\sigma^z_{p-1}\cdots \sigma^z_0$ enforces fermionic anticommutation. Consequently, an one-body/two-body term in the Hamiltonian can be expanded as a linear combination of 4/16 Pauli strings, respectively. In a conventional implementation, one would construct an intermediate fermionic operator and then symbolically apply the Jordan--Wigner mapping. In this work, for each non-negligible integral, the corresponding 4 or 16 Pauli $(X,Z)$ masks and their phase-corrected coefficients are precomputed directly via bitwise operations and accumulated into a hash table keyed by $(X,Z)$, as summarized in Algorithm~\ref{alg:jw_construction}. The same framework can be readily extended to other fermion-to-qubit mappings, such as the parity and Bravyi--Kitaev transformations, by replacing the corresponding Pauli expansion rules. It is worth noting that the present framework is also well suited for constructing and manipulating effective Hamiltonians, where Pauli-structured operator product can lead to a rapid growth in the number of Pauli terms, as encountered in qubit coupled-cluster and effective Hamiltonian approaches. \cite{Ryabinkin2018QCC,Ryabinkin2020iQCC,LiuLiYang2022,Lang2023}

\subsection{Operator--State Multiplication}
\label{sec:operator_state_mul}

In the binary symplectic representation, each Pauli string is encoded using $(X,Z, \Lambda)$. The $\sigma_x$-mask $X$ determines which computational-basis bits are flipped, while the $\sigma_z$-mask $Z$ determines the phase accumulated on a given basis state. As a result, many Pauli strings may induce the same transition between computational-basis configurations, differing only in their parity masks and coefficients.

Let ${X}$ denote the bit-flip signature of a Pauli string. For a computational-basis state $|b_s\rangle$ in the binary representation, all Pauli strings with the same ${X}$ map the state to the same target basis state $|b_s\oplus {X}\rangle$, although they may contribute different phase factors through their corresponding $\sigma_z$-masks. This allows us to separate the bit-flip pattern from the phase-evaluation step. Consequently, one can categorize $\hat{A} = \sum_{k=1}^M c_k \hat{P}_k$ into groups labeled by unique $\sigma_x$-masks,
\begin{equation}
    \hat{A}
    =
    \sum_{g=1}^{G}
    \mathcal{G}_g,
    \qquad
    \mathcal{G}_g
    =
    \mathcal{R}\left(
    {X}_g,
    \mathbf{Z}_g,
    \mathbf{C}_g
    \right),
    \label{eq:topological_group}
\end{equation}
where $G$ is the number of distinct bit-flip patterns. The array $\mathbf{Z}_g=\{Z_k \mid X_k=X_g\}$ collects all $\sigma_z$-masks associated with group $g$, while $\mathbf{C}_g=\{c_k^g \mid X_k=X_g\}$ stores the corresponding coefficients. For a state vector
\begin{equation}
    |\psi\rangle
    =
    \sum_{s=0}^{D-1}
    \psi_{b_s} |b_s\rangle ,
\end{equation}
the action of $\hat{A}$ can then be formulated as
\begin{equation}
    \hat{A}|\psi\rangle
    =
    \sum_{g=1}^{G}
    \sum_{s=0}^{D-1}
    \psi_{b_s}
    \left[
    \sum_{k=1}^{N_g}
    c_k^g
    (-1)^{
    \phi({Z}_k , b_s)
    }
    \right]
    | {X}_g \oplus b_s\rangle ,
    \label{eq:grouped_contraction}
\end{equation}
where $D$ is the dimension of configurations with fixed particle number and spin, and $N_g$ is the number of Pauli strings in group $g$.  Consequently, \textit{$\sigma_x$-mask grouping} reduces the number of distinct configuration-transition and basis-indexing loops from the total number of Pauli strings $M$ to the number of bit-flip groups $G$, while the remaining group-internal summation involves only phase-factor and coefficient accumulation. In practical calculations, $G$ is typically much smaller than $M$ in $\hat{A}$, as reflected by the $M/G$ values listed in Table~\ref{tab:binsim_csf_timing}. This reduction in transition loops provides a direct source of acceleration for operator--state multiplication.

In a source-configuration-based implementation of Eq.~\eqref{eq:topological_group}, different source configurations and $\sigma_x$-mask groups may contribute to the same target configuration. In parallel calculations, this results in simultaneous updates of the same output-vector element, which must be handled by synchronization operations or temporary thread-local buffers. Both choices introduce extra memory traffic and summation overhead. To avoid this bottleneck, we combine $\sigma_x$-mask grouping with target-configuration-based summation. Because all Pauli strings in the same group $g$ share an identical bit-flip mask $X_g$, the source configuration contributing to a fixed target configuration $|b_t\rangle$, is uniquely determined as $|b_s\rangle=|b_t \oplus X_g\rangle$. Therefore, instead of scattering contributions from each source configuration to multiple targets, we iterate over target configurations and gather all compatible source amplitudes, namely,
\begin{equation}
    \hat{A}|\psi\rangle
    =
    \sum_{t=0}^{D-1}
    \left[\sum_{g=1}^{G}
    \psi_{b_t \oplus X_g}
    \left(\sum_{k=1}^{N_g}
    c_k^g
    (-1)^{
    \phi(b_{t}\oplus X_{g} ,Z_{k})
    } \right)\right]
    |b_{t}\rangle .
    \label{eq:reverse_sum}
\end{equation}
It is clear that the innermost summation of Eq.~\eqref{eq:reverse_sum} evaluates the phase interference among all Pauli strings sharing the same bit-flip mask. The middle loop accumulates the coefficient of $|b_t\rangle$ from the source configuration $|b_s\rangle = | b_t \oplus {X}_g \rangle$. The outermost loop iterates over all target configurations. In this form, each target-configuration coefficient is evaluated independently, so the summation over $t$ can be parallelized directly without simultaneous updates to the same output-vector element. 

In addition, the $n$-qubit computational basis can be partitioned into a spin-resolved product basis, $ |b_s\rangle = |b_{s_{\alpha}} \otimes b_{s_{\beta}}\rangle$. Under this decomposition, the bit masks are decomposed as
${X}_g=({X}_{g_{\alpha}},{X}_{g_{\beta}})$ and
$Z_k=(Z_{k_{\alpha}},Z_{k_{\beta}})$, and the target configuration becomes
\begin{equation}
    |b_s \oplus X_{g}\rangle 
    =
    |b_{s_{\alpha}}\oplus X_{g_{\alpha}}\rangle
    \otimes
    |b_{s_{\beta}}\oplus X_{g_{\beta}}\rangle .
\end{equation} 
Next, considering molecular point-group symmetry, the wave function in the target symmetry sector can be expanded as
\begin{equation}
    |\psi\rangle
    =
    \sum_r \sum_{t_\alpha=1}^{D^r_\alpha} \sum_{t_\beta=1}^{D^r_\beta}
    \psi_{b_t^r}
    |b_{t_\alpha}^r \otimes b_{t_\beta}^r\rangle, 
\end{equation}
where $r$ labels an allowed pair of spin-string irreducible representations
$(\Gamma_\alpha^r,\Gamma_\beta^r)$. A target configuration in block $r$ is denoted as
\begin{equation}
    |b_t^r\rangle
    =
    |b_{t_\alpha}^{r}
    \otimes
    b_{t_\beta}^{r}\rangle,
    \qquad
    \Gamma(b_{t_\alpha}^{r})\otimes
    \Gamma(b_{t_\beta}^{r})
    =
    \Gamma(b_{t}^r).
\end{equation}
Here, $\Gamma(\cdot)$ denotes the irreducible representation of a spin bitstring. As such, Eq.~\eqref{eq:reverse_sum} can be rewritten as 
\begin{align}
    \hat{A}|\psi\rangle
    =
    \sum_r
    \sum_{t_{\alpha}=0}^{D^r_\alpha-1}
    \sum_{t_{\beta}=0}^{D^r_\beta-1}
    \left[\sum_{g=1}^{G}
    \psi_{b^r_t \oplus X_g}
    \sum_{k=1}^{N_g}
    c_k^g
    (-1)^{
    \phi(b^r_{t_{\alpha}}\oplus X_{g_\alpha} ,Z_{k_{\alpha}})
    }\right. \nonumber \\ 
    \left.
    (-1)^{
    \phi(b^r_{t_{\beta}}\oplus X_{g_\beta},Z_{k_{\beta}})
    }
    \right]
    |b^r_{t_\alpha} \otimes b^r_{t_\beta}\rangle .
    \label{eq:sreverse_sum}
\end{align}
Here, the source configuration $|b_t^r\oplus X_g\rangle$ is resolved separately in the $\alpha$ and $\beta$ sectors, which replaces a direct lookup in the full determinant space by two smaller spin-string index searches. The phase factor also factorizes into independent $\alpha$ and $\beta$ contributions, enabling sign evaluations to be carried out in smaller spin-string spaces. In addition, the symmetry-block organization leads to more regular access to CI-vector amplitudes and improves data locality during operator--state multiplication. This symmetry-adapted form is particularly beneficial for large-scale state-vector simulations, where memory bandwidth and cache efficiency often dominate the overall performance.\cite{smelyanskiy2016qhipster,jones2019quest} It is worth noting that the target-configuration-based summation and symmetry blocking have been widely used in conventional full configuration interaction algorithms.\cite{knowles1984new,olsen1988determinant} Here, we reformulate these techniques for Pauli-algebra operations.

However, when directly evaluating the phase sums in Eq.~\eqref{eq:sreverse_sum}, each $(t_\alpha, t_\beta)$ pair requires $\mathcal{O}(N_g)$ bitwise parity checks, incurring both a popcnt latency and a data-dependent conditional negation for each check. In the following, we show that a tensor rank decomposition (SVD) of each group's phase matrix can eliminate these bitwise operations entirely from the hot loop, reducing the innermost contraction to pure fused multiply--add (FMA) vector operations.

To this end, we note that the $(Z_{k_\alpha}, Z_{k_\beta})$ pairs are often linearly dependent across different $k$, making it possible to factor the parity evaluation into a spin-resolved product via singular value decomposition (SVD). We first extract the unique $\alpha$ and $\beta$ parts of the $Z$-masks,
\begin{equation}
\mathbf{\hat{Z}}_g^\alpha = \{\hat{Z}_\mu^\alpha\}_{\mu=1}^{N_g\alpha},
\qquad
\mathbf{\hat{Z}}_g^\beta = \{\hat{Z}_\nu^\beta\}_{\nu=1}^{N_g^\beta},
\end{equation}
and assemble the coefficient matrix $\mathbf{M}^{(g)} \in \mathbb{R}^{N_g^\alpha \times N_g^\beta}$, where $M^{(g)}_{\mu\nu}$ is the sum of all $c_k^g$ whose $Z$-mask equals $(\hat{Z}_\mu^\alpha, \hat{Z}_\nu^\beta)$. Decomposing $\mathbf{M}^{(g)}$ via $\mathbf{M}^{(g)} = \mathbf{U}\mathbf{\Sigma}\mathbf{V}^T$ yields  
\begin{equation}
M^{(g)}_{\mu\nu} = \sum_{k=1}^{R_g} U_{\mu k}\,\sigma_k\, V_{\nu k}
\quad\Longrightarrow\quad
\begin{cases}
W_{\mu k}^{(g_\alpha)} = U_{\mu k}\sqrt{\sigma_k},\\[4pt]
W_{\nu k}^{(g_\beta)}  = V_{\nu k}\sqrt{\sigma_k}.
\end{cases}
\label{eq:svd_weights}
\end{equation}
where $R_g$ is the matrix rank of $\mathbf{M}^{(g)}$.

Substituting the coefficients $c_k^g$ with the rank-$R_g$ expansion defined by $W$, the phase evaluation naturally factorizes into independent spin components:
\begin{align}
F_\alpha^g(t_\alpha, k) &= \sum_{\mu=1}^{N_g^\alpha}
W_{\mu k}^{(g_\alpha)}\,
(-1)^{\phi(b^r_{t_{\alpha}}\oplus X_{g_\alpha} ,Z_{k_{\alpha}})},
\label{eq:precomp_alpha} \\
F_\beta^g(t_\beta, k) &= \sum_{\nu=1}^{N_g^\beta}
W_{\nu k}^{(g_\beta)}\,
(-1)^{\phi(b^r_{t_{\beta}}\oplus X_{g_\beta},Z_{k_{\beta}})},
\label{eq:precomp_beta}
\end{align}

With this factorization, the original phase sums in the inner loop reduce to a rank-$R_g$ dot product of the precomputed scalars $F_\alpha^g$ and $F_\beta^g$:
\begin{align}
    \hat{A}|\psi\rangle
    =
    \sum_r
    \sum_{t_{\alpha}=0}^{D^r_\alpha-1}
    \sum_{t_{\beta}=0}^{D^r_\beta-1}
    \left[\sum_{g=1}^{G}
    \psi_{b^r_t \oplus X_g}
    \sum_{k=1}^{R_g}
    F_\alpha^g(t_\alpha, k) \cdot F_\beta^g(t_\beta, k)
    \right]
    |b^r_{t_\alpha} \otimes b^r_{t_\beta}\rangle .
    \label{eq:svd_sreverse_sum}
\end{align}

For the two-body electronic Hamiltonian under the Jordan--Wigner mapping, the matrix $\mathbf{M}^{(g)}$ exhibits a near-universal rank collapse: off-diagonal groups ($X_g \neq 0$) satisfy $R_g \le 2$ regardless of the number of accumulated Pauli terms $N_g$, while the diagonal group ($X_g = 0$) has $R_0 = n_{\rm orb} + 2$, where the $+2$ originates from the identity operator and the pure one-body diagonal, and the remaining rank $n_{\rm orb}$ reflects the cross-spin Coulomb matrix $\sum_{pq} V_{pq} \hat{n}_{p,\alpha}\hat{n}_{q,\beta}$. Table~\ref{tab:svd_rank_dist} confirms this regularity across eleven molecular systems spanning 18--29 spatial orbitals and three basis-set tiers.

\begin{table}[!htb]
\centering
\small
\setlength{\tabcolsep}{10pt}
\begin{tabular}{lcccccccc}
\toprule
System      & Basis     & $n_{\rm orb}$ & $G$       & $R_g=1$       & $R_g=2$       &$\overline{R}$ \\
\midrule
C$_2$       & 6-31G     & 18            &  4266     & 98.9\%        & 1.1\%         & 1.011         \\
N$_2$       & 6-31G     & 18            &  4266     & 98.9\%        & 1.1\%         & 1.011         \\
CO          & 6-31G     & 18            &  8396     & 98.6\%        & 1.4\%         & 1.013         \\
H$_2$O      & cc-pVDZ   & 24            & 25495     & 99.3\%        & 0.7\%         & 1.006         \\
O$_2$       & 6-31G     & 18            &  4266     & 98.9\%        & 1.1\%         & 1.011         \\
C$_3$H$_4$  & STO-3G    & 19            & 13976     & 98.4\%        & 1.6\%         & 1.016         \\
HCN         & 6-31G     & 20            & 13755     & 98.9\%        & 1.1\%         & 1.011         \\
C$_2$H$_2$  & 6-31G     & 22            & 10908     & 99.1\%        & 0.8\%         & 1.008         \\
NH$_3$      & cc-pVDZ   & 29            & 92711     & 99.5\%        & 0.5\%         & 1.005         \\
C$_2$       & cc-pVDZ   & 28            & 21783     & 99.5\%        & 0.5\%         & 1.005         \\
C$_3$H$_6$  & STO-3G    & 21            & 31467     & 99.1\%        & 0.9\%         & 1.009         \\
\bottomrule
\end{tabular}
\caption{Distribution of the SVD rank $R_g$ of the phase matrix $\mathbf{M}^{(g)}$ for different molecular systems. Off-diagonal groups are exhausted by $R_g = 1$ and $R_g = 2$. All calculations used a truncation tolerance $\sigma_r > 10^{-12}$.}
\label{tab:svd_rank_dist}
\end{table}

Equations~\eqref{eq:precomp_alpha}--\eqref{eq:precomp_beta} contain all popcnt evaluations, but they lie \textit{outside} the $(t_\alpha,t_\beta)$ double loop. The per-$(t_\alpha, t_\beta)$ cost then reduces to the dot product:
\begin{equation}\label{eq:dot_product}
    f_g = \sum_{k=1}^{R_g}
    F_\alpha^g(t_\alpha, k) \cdot F_\beta^g(t_\beta, k)
\end{equation}
which involves only $R_g$ contiguous fused multiply--add operations — no popcnt, no conditional negation, and fully vectorizable via SIMD. Details for implementing Eq.~\eqref{eq:svd_sreverse_sum} are summarized in Algorithm~\ref{alg:grouped_hvec}.

A naive implementation that precomputes and stores all $F_\beta^g$ arrays before the contraction would require storage proportional to $G \cdot D_\beta^r \cdot\overline{R}$, which for large basis sets can exceed the wavefunction memory itself because the number of groups $G$ grows rapidly with $n_{\rm orb}$. We avoid this overhead by computing the phase vectors on the fly (OTF) in a batched scanning mode: at runtime, groups are processed in small batches of size $B$; the phase vectors are precomputed only for the current batch, consumed immediately in the $(t_\alpha,t_\beta)$ hot loop, and then discarded. The peak auxiliary memory is thus $B \cdot D_\beta^r \cdot \overline{R}$, which is strictly bounded by a small multiple of the wavefunction vector size, making the memory footprint independent of $G$.

Overall, SVD compression yields two complementary sources of acceleration: (i) roughly a factor of $\sim\!5$ reduction in the total number of accumulated terms, as reflected by the $M/G$ values listed in Table~\ref{tab:binsim_csf_timing}, and (ii) transformation of the inner-loop arithmetic from latency-bound popcnt and branch evaluation to throughput-bound SIMD FMAs. In practice, the combined effect yields a 50$\times$+ speedup for the Hamiltonian--vector multiplication kernel. The same SVD-based grouped evaluation applies to excitation operators and other Pauli-structured operator pools. For fermionic operators under standard qubit mappings---such as those appearing in UCC-type ansatze---the coefficient matrices $\mathbf{M}^{(g)}$ inherit the low-rank structure analyzed above; for more general operators, the SVD still provides significant compression in practice.

\begin{algorithm}[!htb]
\SetKwData{Xor}{\ensuremath{\mathbin{\oplus}}}

\DontPrintSemicolon
\SetKwFunction{GroupedHvec}{Hvec}
\SetKwFunction{Precompute}{Precompute}
\SetKwFunction{DotProduct}{DotProduct}
\SetKw{Function}{Function}
\Function{\GroupedHvec{$\hat{A}$, $|\psi \rangle$}}\;
\KwIn{SVD-compressed operator $\hat{A}$, state vector $|\psi \rangle$}
\KwOut{$|\psi'\rangle = \hat{A}\,|\psi \rangle$}
\PrintSemicolon

$|\psi' \rangle \gets 0$ \;
\For{each \rm{target block} $r$}{
    \For{$j \gets 0$ \KwTo $D^{r}_\beta - 1$}{
        \For{each \rm{group} $g$}{
            $F_\beta^g[j] \gets (g, b^r_{t_\beta} \oplus X_{g_\beta})$ \tcp*{Eq.\eqref{eq:precomp_beta}}
        }
    }
    \For{$i \gets 0$ \KwTo $D^{r}_\alpha - 1$}{
        \For{each \rm{group} $g$}{
            $F_\alpha^g \gets (g, b^r_{t_\alpha} \oplus X_{g_\alpha})$ \tcp*{Eq.\eqref{eq:precomp_alpha}}
            $b_{s_\alpha} \gets b^r_{t_\alpha} \oplus X_{g_\alpha}$\;
            \For{$j \gets 0$ \KwTo $D^{r}_\beta - 1$}{
                $b_{s_\beta} \gets b^r_{t_\beta} \oplus X_{g_\beta}$\;
                $f \gets (F_\alpha^g, F_\beta^g[j])$ \tcp*{Eq.\eqref{eq:dot_product}}
                $\psi'_{b^r_t} \gets \psi'_{b^r_t} + \psi_{b_s} \cdot f$\;
            }
        }
    }
}
\Return{$|\psi'\rangle$}
\caption{Grouped operator--state multiplication with target-driven gather and tensor-rank factorization (SVD).}
\label{alg:grouped_hvec}
\end{algorithm}

\subsection{Universality of the Grouped Framework} 

It is important to emphasize that the present framework is built on generalized Pauli algebra rather than on problem-specific physical symmetries, and is therefore largely agnostic to the physical origin of the operator. Conventional simulation solvers are often optimized for particular conserved quantum numbers, such as particle number or spin projection, which can lead to reduced flexibility when these symmetries are absent or explicitly broken. In contrast, the proposed bit-flip-based routing only depends on the binary symplectic structure of the Pauli strings and naturally accommodates arbitrary Pauli operators. As a result, the framework can be applied not only to standard particle-number-conserving electronic Hamiltonians, but also to particle-number-nonconserving models, such as Bardeen--Cooper--Schrieffer superconducting Hamiltonians,\cite{wu2002polynomial} frustrated quantum spin lattices,\cite{bassman2021simulating} and general non-physical Pauli operator pools used in variational or adaptive quantum algorithms.

Furthermore, because the operator action is decomposed into independent groups labeled by distinct bit-flip patterns, the framework is well suited for step-wise state evolution. Examples include the sequential application of Pauli exponentials, fermionic excitation operators, or exponentiated unitary-cluster generators in variational ans\"atze. In such cases, the grouped representation allows memory routing and phase accumulation to be reused within each group, providing a flexible and efficient algebraic backend for large-scale emulation of parameterized quantum circuits.

\section{Results and Discussion}
\label{sec:results}

\subsection{Pauli-Algebra Benchmarks}
\subsubsection{Jordan--Wigner Transformation}

We first assess the performance of the proposed Pauli-algebra algorithm for constructing qubit Hamiltonians from electronic integrals via the Jordan--Wigner transformation. The implementation developed in this work, referred to as BinSim, is integrated into the Q$^2$Chemistry package.\cite{JUSTC-2022-0118} Benchmark calculations were performed using randomly generated, fully dense one- and two-electron integrals in double precision. The one-electron integral matrices were constrained to be Hermitian, whereas the two-electron integrals satisfied the standard eightfold permutation symmetry. For comparison, the electronic Hamiltonian in OpenFermion-1.7.0 was first constructed as an \texttt{InteractionOperator} and subsequently transformed into its qubit representation using the built-in \texttt{jordan\_wigner} routine.\cite{mcclean2020openfermion} BinSim was executed using 1, 8, and 64 CPU threads, denoted as BinSim-1t, BinSim-8t, and BinSim-64t, respectively.

\begin{figure}[!htb]
\centering
\includegraphics[width=0.95\linewidth]{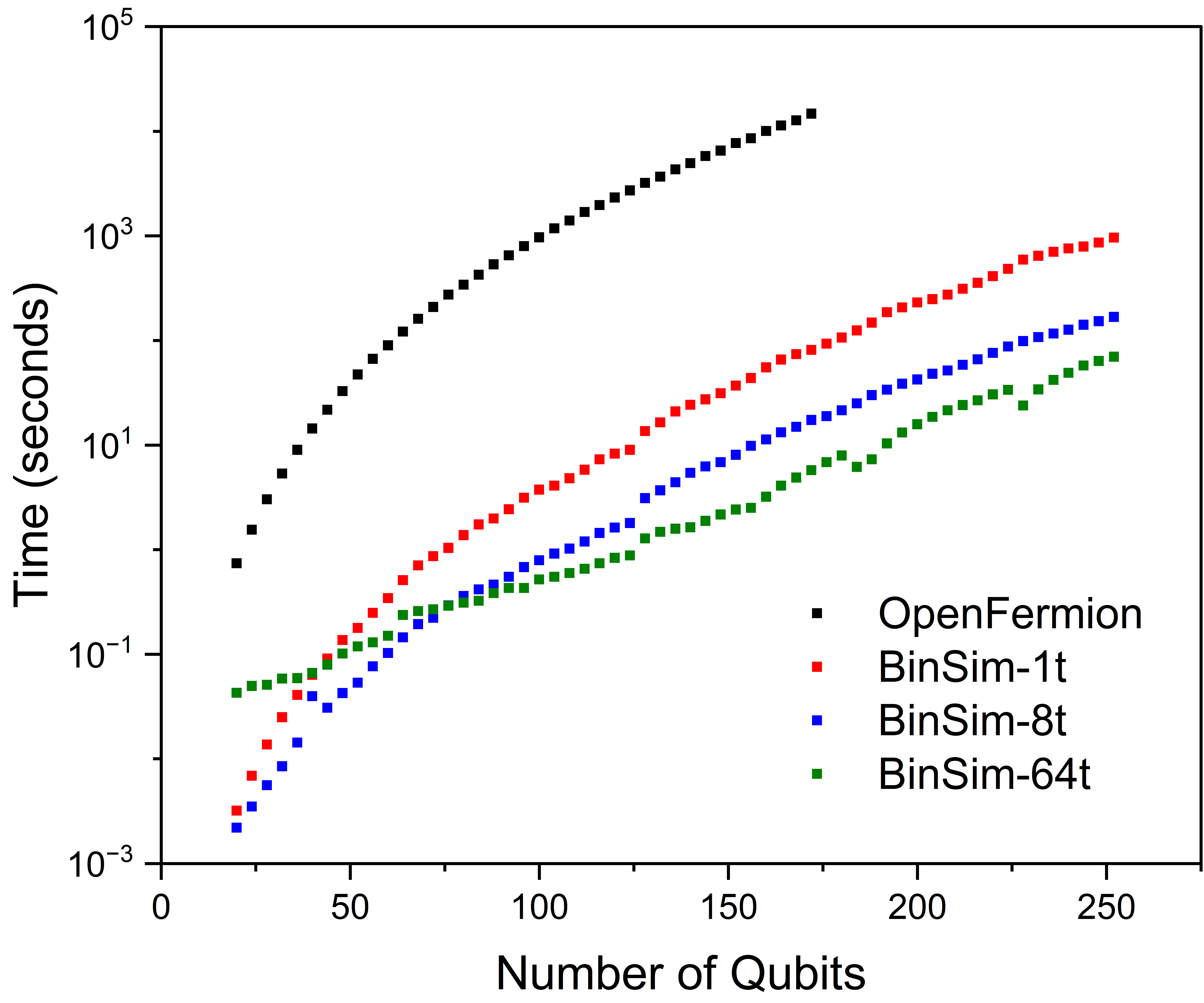}
\caption{Wall-clock time required for the full Jordan--Wigner transformation as a function of the number of qubits. OpenFermion was executed serially using \texttt{InteractionOperator} and \texttt{jordan\_wigner}, while BinSim was executed with 1, 8, and 64 CPU threads, labeled BinSim-1t, BinSim-8t, and BinSim-64t, respectively. }
\label{fig:jw_benchmark}
\end{figure}

As shown in Fig.~\ref{fig:jw_benchmark}, even the single-threaded BinSim implementation substantially outperforms serial OpenFermion. This single-thread comparison eliminates differences arising from parallel scaling and therefore provides a more direct assessment of the underlying algorithmic performance. The speedup, defined as $t_{\rm OpenFermion}/t_{\rm BinSim\text{-}1t}$, ranges from $171\times$ to $302\times$ over the tested system sizes, demonstrating that BinSim maintains a substantial computational advantage throughout the benchmark.

In addition, BinSim demonstrates effective parallel scaling. Relative to BinSim-1t, both BinSim-8t and BinSim-64t achieve increasingly substantial speedups as the number of qubits grows. The parallel benefit is less pronounced for smaller systems, for which thread-management overhead and parallel reduction costs account for a larger fraction of the total runtime. As the Hamiltonian size increases, the rapidly growing number of generated Pauli terms provides sufficient computational workload to expose greater parallelism, making BinSim-64t the fastest among the tested BinSim configurations. Moreover, whereas the serial OpenFermion calculations become quite expensive beyond the largest systems included in the direct comparison, BinSim remains capable of completing the Jordan--Wigner transformation for systems containing up to 252 qubits.

\subsubsection{Operator-State Multiplication}

All benchmarks of operator-state multiplication were performed on a single CPU node equipped with two AMD EPYC-9A14 processors, comprising 192 physical CPU cores and 768 GB of memory. Hartree--Fock calculations were carried out using PySCF-2.8.0,\cite{sun2018pyscf} which was also used to generate the one- and two-electron integrals. Unless otherwise specified, all reported timings correspond to wall-clock times averaged over multiple independent runs.

The numerical benchmarks were designed to cover the major Pauli-algebra workloads encountered in quantum-chemistry and many-body simulations, including fermion-to-qubit transformations, the application of Trotterized unitary coupled-cluster operators, and Hamiltonian--vector multiplication. These operations constitute key computational hotspots in both static and dynamical simulations.
For molecular calculations, the notation $(N_ee,N_oo)$ denotes an active space with $N_e$ electrons in $N_o$ spatial orbitals, corresponding to $2N_o$ spin orbitals for Jordan--Wigner transformation.\cite{jordan1928pauli}

Table~\ref{tab:binsim_csf_timing} summarizes the molecular systems used for Pauli-algebra benchmarks. The column ``Det'' labels the dimension of configurations with fixed particle number and spin,
\begin{equation} \label{eq:fcidim}
    {\rm Det} = \dim \mathcal{H}
    =
    \binom{N_o}{N_\alpha}
    \binom{N_o}{N_\beta},
\end{equation}
whereas ``sDet'' denotes the number of configuration in the target symmetry of the ground state according to the Abelian point-group symmetry specified in the ``Symm'' column. The latter is the actual working space used by BinSim in the VQE-related benchmarks. The ``Trotter'' column reports the wall-clock time required to apply the first-order Trotterized unitary coupled cluster (UCC) with singles and doubles (UCCSD) ansatz to the state vector,
\begin{equation}
    |\Psi(\boldsymbol{\theta})\rangle
    =
    \prod_k \exp(\theta_k \hat{\tau}_k)
    |\Phi_{\rm HF}\rangle ,
\end{equation}
where $\hat{\tau}_k$ are anti-Hermitian single and double excitation generators.\cite{McAEndAsp20,Tilly_VQE_2021} The ``Hv'' column reports the wall-clock time required to apply the molecular Hamiltonian to a random state vector, while the ``E'' column reports the time required for a single energy evaluation, $\langle \Psi|\hat{H}|\Psi\rangle$, corresponding to an energy-only VQE iteration without gradient evaluation.

\begin{table}[H]
\centering
\small
\setlength{\tabcolsep}{5pt}
\begin{tabular}{cccccccccc}
\hline
System      & Basis   & Size       & Symm       & Det                   & sDet                      & $M/G$          & Trotter       & Hv            & E     \\
\hline
C$_2$       & 6-31G   & (12e, 18o) & D$_{2h}$   & 3.4$\times10^{8}$     & 4.3$\times10^{7}$         & 5.28         & 0.5           & 1.1           & 1.5   \\
N$_2$       & 6-31G   & (14e, 18o) & D$_{2h}$   & 1.0$\times10^{9}$     & 1.3$\times10^{8}$         & 5.28         & 1.6           & 2.7           & 4.3   \\
CO          & 6-31G   & (14e, 18o) & C$_{2v}$   & 1.0$\times10^{9}$     & 2.5$\times10^{8}$         & 5.35         & 6.6           & 9.7           & 18.6  \\
H$_2$O      & cc-pVDZ & (10e, 24o) & C$_{2v}$   & 1.8$\times10^{9}$     & 4.5$\times10^{8}$         & 5.05         & 12.5          & 28.2          & 43.1  \\
O$_2$       & 6-31G   & (16e, 18o) & D$_{2h}$   & 1.9$\times10^{9}$     & 2.4$\times10^{8}$         & 5.28         & 5.5           & 5.2           & 10.5  \\
C$_3$H$_4$  & STO-3G  & (22e, 19o) & C$_s$      & 5.7$\times10^{9}$     & 2.9$\times10^{9}$         & 5.22         & 234           & 239           & 450   \\
HCN         & 6-31G   & (14e, 20o) & C$_{2v}$   & 6.0$\times10^{9}$     & 1.5$\times10^{9}$         & 5.28         & 71.7          & 95.9          & 174   \\
C$_2$H$_2$  & 6-31G   & (14e, 22o) & D$_{2h}$   & 2.9$\times10^{10}$    & 3.6$\times10^{9}$         & 5.18         & 122           & 208           & 331   \\
NH$_3$      & cc-pVDZ & (10e, 29o) & C$_s$      & 1.4$\times10^{10}$    & 7.1$\times10^{9}$         & 4.84         & 1725          & 2090          & 3592  \\
C$_2$       & cc-pVDZ & (12e, 28o) & D$_{2h}$   & 1.4$\times10^{11}$    & 1.8$\times10^{10}$        & 4.95         & 578           & 1917          & 2286  \\
C$_3$H$_6$  & STO-3G  & (24e, 21o) & C$_s$      & 8.6$\times10^{10}$    & 4.3$\times10^{10}$        & 5.14         & 5625          & 15180         & 22711 \\
\hline
\end{tabular}
\caption{Pauli-algebra benchmarks. ``Symm'' denotes the Abelian point group used in the calculations. ``Det'' and ``sDet'' gives the dimension of configurations before and after point-group reduction, respectively. $M/G$ denotes the average number of Pauli strings per $\sigma_x$-mask group in the Hamiltonian. ``Trotter'' reports the time required to apply the first-order Trotterized UCCSD ansatz with random parameters to the state vector once. ``Hv'' reports the Hamiltonian--vector multiplication time for a random state vector, and ``E'' reports the time required for a single variational energy evaluation, $\langle\Psi|\hat{H}|\Psi\rangle$, corresponding to an energy-only VQE iteration without gradient evaluation. All timings are wall-clock times in seconds obtained on a single dual EPYC-9A14 CPU node and averaged over repeated runs.}
\label{tab:binsim_csf_timing}
\end{table}

As shown in table~\ref{tab:binsim_csf_timing}, the computational cost is not determined solely by the number of active orbitals, but also depends strongly on the dimension of symmetry-reduced full configuration interaction (FCI) space and on the sparsity pattern of the corresponding operator action. For relatively small benchmarks, such as C$_2$/6-31G, N$_2$/6-31G, O$_2$/6-31G, and CO/6-31G, the full FCI spaces already contain $10^8$--$10^9$ configurations, yet a variational energy evaluation can still be completed within seconds. As the symmetry-reduced FCI space grows, the computational timings increase to the minute scale for HCN, C$_2$H$_2$, and C$_3$H$_4$, and eventually reach the hour scale for the largest C$_3$H$_6$ benchmark. The comparison between systems with similar active-space sizes further indicates that symmetry reduction and operator sparsity play a decisive role: for example, NH$_3$/cc-pVDZ has a smaller symmetry-reduced FCI dimension than C$_2$/cc-pVDZ, but exhibits a longer energy-evaluation time, reflecting differences in the effective connectivity of the Hamiltonian and UCCSD operator actions. The ``E'' timings are generally comparable to the combined cost of state preparation through the Trotterized UCCSD ansatz and one Hamiltonian--vector multiplication, confirming that these two kernels dominate an energy-only VQE iteration. Overall, these benchmarks demonstrate that Binsim provides not only compact storage of large molecular operators, but also an efficient module for applying Hamiltonians and UCC-type operators directly in large symmetry-adapted many-electron spaces.

\subsubsection{Comparison with Other Simulators}

To contextualize the performance of BinSim, we compare it against three representative simulators---the dense circuit simulator Qiskit-aer~2.5.0, and the sparse fermionic simulators ffsim~0.0.83 and fqe~0.3.0---on C$_2$H$_4$/STO-3G with an active space of (16e, 14o). Table~\ref{tab:sim_comparison} reports wall-clock times for Trotterized UCCSD application and Hamiltonian--vector (Hv) multiplication on a single CPU node with~1 and~8 OpenMP threads. The nearly six-orders-of-magnitude gap between BinSim and Qiskit-aer arises from the inherent cost of dense circuit simulation: applying $\sim$10$^5$ gates to a $2^{28}$-dimensional statevector is prohibitively expensive, and this workload falls squarely in a regime where circuit simulators become impractical. In contrast, against the sparse fermionic simulators, BinSim achieves a $\sim$265$\times$ speedup for the multi-threaded Trotterized UCCSD application, $\sim$50$\times$ speedup for the multi-threaded Hv operation. Per-simulator backend configurations and version details are provided in the Appendix.

\begin{table}[!htb]
\centering
\caption{Wall-clock time (seconds) for Trotterized UCCSD application (Trotter) and Hamiltonian--vector multiplication (Hv) on C$_2$H$_4$/STO-3G (16e,14o). Qiskit-aer Hv is not available; ffsim Trotter does not support multi-threading.}
\label{tab:sim_comparison}
\begin{tabular}{l c r r}
\toprule
Simulator   & Threads & Trotter  & Hv \\
\midrule
Qiskit-aer  & 1  & 117\,847 & -- \\
ffsim       & 1  & 38.8     & 21.6 \\
fqe         & 1  & 15.3     & 18.1 \\
BinSim      & 1  & 0.20     & 0.56 \\
\midrule
Qiskit-aer  & 8  & 18\,071  & -- \\
ffsim       & 8  & --       & 2.90 \\
fqe         & 8  & 10.6     & 4.27 \\
BinSim      & 8  & 0.04     & 0.08 \\
\bottomrule
\end{tabular}
\end{table}

\subsection{VQE Simulations of Molecular Dissociation}\label{ssec:PES}

Here, we used BinSim as the backend for VQE simulations of molecular dissociation potential energy surfaces (PESs) based on Trotterized unitary coupled-cluster (UCC) ans\"atze. In these calculations, the trial state was prepared from the Hartree--Fock determinant using either a first-order Trotterized UCCSD, denoted as tUCCSD, or a first-order Trotterized UCC with generalized single and double (GSD) excitations, denoted as tUCCGSD. The variational parameters were optimized with the L-BFGS algorithm implemented in the Julia Optim package. All line searches were performed using the More--Thuente algorithm.~\cite{MoreThuente1994Linesearch} In these calculations, tUCCSD-VQE can be readily optimized using tight convergence thresholds of $g_{\rm tol}=10^{-6}$ and $f_{\rm tol}=10^{-8}$. By contrast, tUCCGSD-VQE involves substantially more variational parameters and is therefore optimized using the less stringent convergence criterion
$(\langle \hat{H}^2\rangle-\langle \hat{H}\rangle^2)/
{\langle \hat{H}\rangle^2}<10^{-7}$.
Despite this looser convergence criterion, tUCCGSD-VQE still achieves an accuracy of 1 kcal/mol with respect to the FCI energy

\begin{figure}[!htb]
\centering
\includegraphics[width=0.7\linewidth]{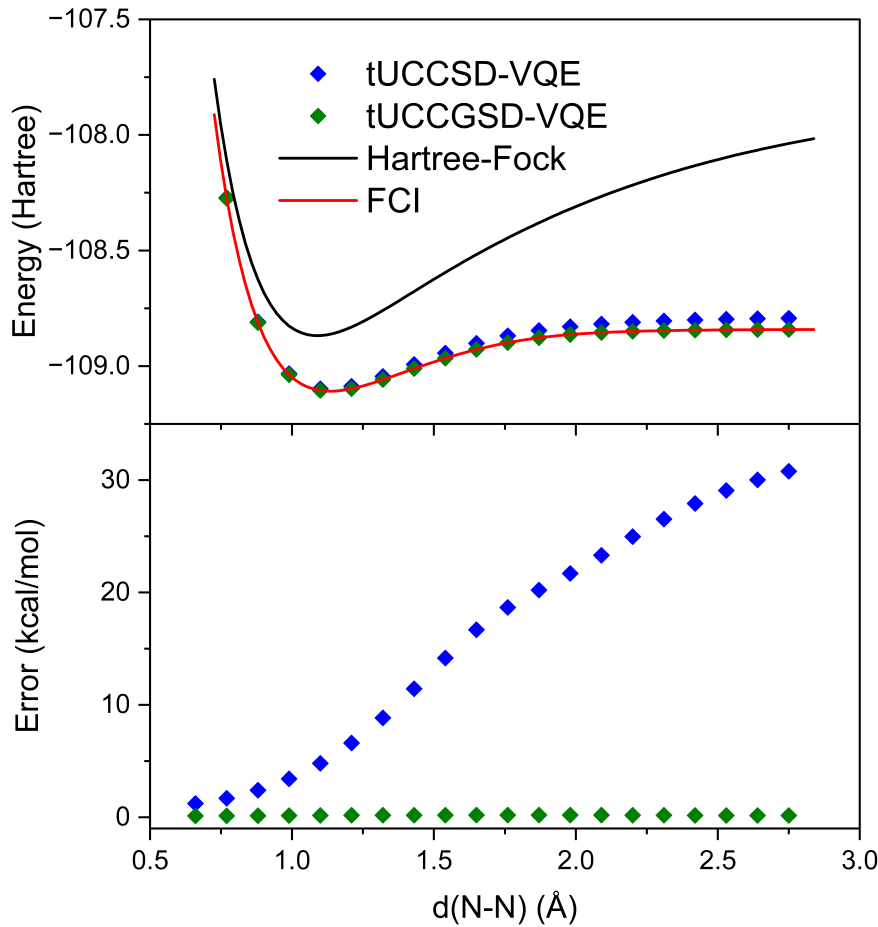}
\caption{Dissociation curve of N$_2$ in the 6-31G basis with active space $(14e,18o)$. In the upper panel, blue diamonds and green diamonds denote the energies obtained from tUCCSD-VQE and tUCCGSD-VQE, respectively. The black and red curves are the Hartree--Fock and FCI energies, both computed with PySCF. In the lower panel, the corresponding absolute energy errors relative to FCI are shown in kcal/mol.}
\label{fig:n2_631g_uccsd}
\end{figure}

Figure~\ref{fig:n2_631g_uccsd} presents the dissociation curve of N$_2$ in the 6-31G basis with an active space of $(14e,18o)$. The Hartree--Fock and FCI reference energies were computed using PySCF. The figure compares the total energies obtained with Hartree--Fock, FCI, tUCCSD-VQE, and tUCCGSD-VQE, together with the absolute errors of the two variational ans\"atze relative to FCI. Both tUCCSD-VQE and tUCCGSD-VQE substantially improve upon Hartree--Fock over the entire bond-length range, demonstrating that the first-order Trotterized coupled-cluster product form captures a significant portion of the electron-correlation energy in both the equilibrium and bond-stretched regions. Nevertheless, the error of tUCCSD-VQE increases rapidly as the N--N bond is stretched, reaching approximately $31$ kcal/mol at a bond length of $2.75$~\AA. This deterioration indicates that the tUCCSD ansatz lacks sufficient flexibility to describe the strong static correlation arising in the dissociation regime, primarily because higher-rank excitation operators are absent.

In contrast, the tUCCGSD-VQE curve remains nearly indistinguishable from the FCI reference over the entire bond-length range. As shown in the lower panel, its absolute error remains below $0.2$ kcal/mol at all geometries, placing it well within the commonly accepted threshold of chemical accuracy. This substantial improvement over tUCCSD-VQE can be understood in terms of the generalized singles-and-doubles excitation manifold, which is not restricted by the occupied--virtual partition defined by the Hartree--Fock reference and therefore provides greater flexibility for describing multireference electronic structures. The comparison between tUCCSD-VQE and tUCCGSD-VQE demonstrates that BinSim can efficiently support both conventional and generalized Trotterized coupled-cluster ansatz, with the latter being essential for achieving quantitatively accurate energies throughout the N$_2$ dissociation process.

The C$_2$ molecule provides a more demanding test because its ground state exhibits substantially stronger multireference character than that of N$_2$. While tUCCSD-VQE accurately captures the predominantly dynamic correlation in N$_2$ near equilibrium, the stronger static correlation and near-degenerate valence configurations in C$_2$ place much higher demands on the ansatz. Consequently, C$_2$ serves as a more stringent benchmark for assessing the capability and limitations of a single-reference tUCCSD ansatz across the dissociation process.

\begin{figure}[!htb]
\centering
\includegraphics[width=0.7\linewidth]{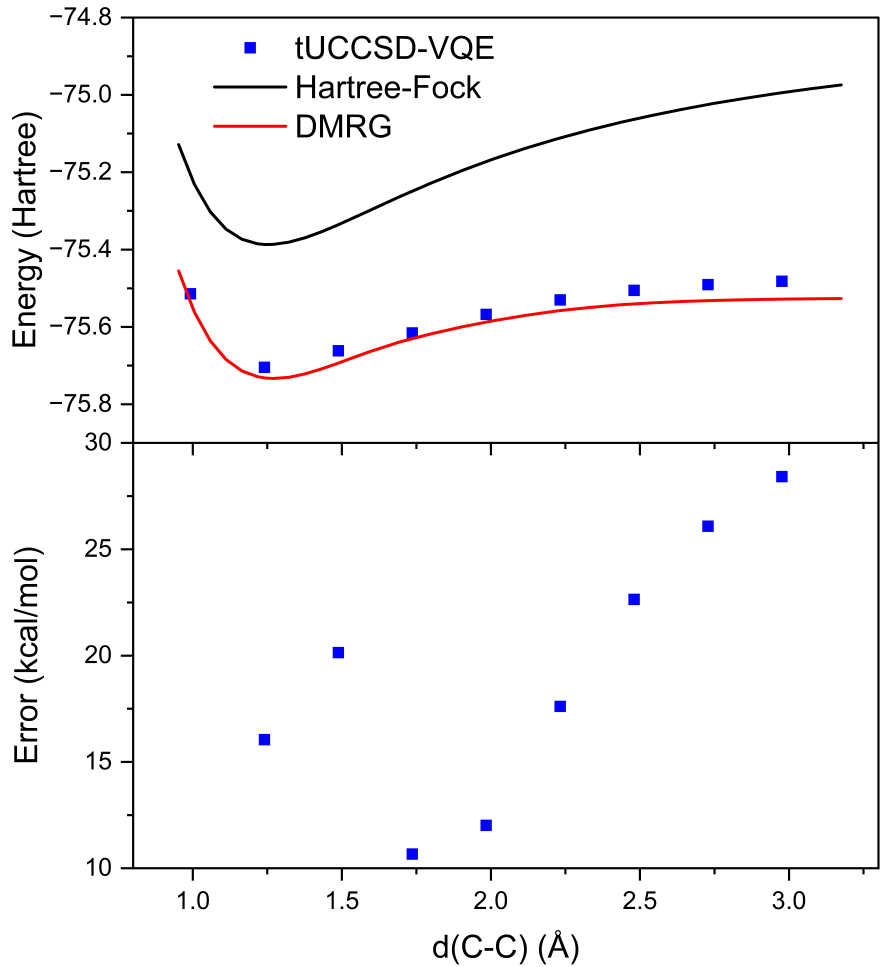}
\caption{Dissociation curve of C$_2$ in the cc-pVDZ basis with active space $(12e,28o)$. In the upper panel, blue squares denote the energies obtained from tUCCSD-VQE. The black curve shows the Hartree--Fock reference, while the red curve shows the DMRG reference for the $X\,{}^1\Sigma_g^+$ state from Ref.~\citenum{Wouters2014CheMPS2}. In the lower panel, the absolute energy error of tUCCSD-VQE relative to DMRG is shown in kcal/mol.}
\label{fig:c2_ccpvdz_uccsd}
\end{figure}

Figure~\ref{fig:c2_ccpvdz_uccsd} shows the C$_2$ dissociation curve in the cc-pVDZ basis with an active space of $(12e,28o)$. The upper panel compares the Hartree--Fock, DMRG, and tUCCSD-VQE energies, while the lower panel reports the absolute tUCCSD-VQE error relative to DMRG in kcal/mol. The Hartree--Fock curve was computed with PySCF, and the DMRG reference curve for the $X\,{}^1\Sigma_g^+$ state was taken from Ref.~\citenum{Wouters2014CheMPS2}. The tUCCSD-VQE calculation used the same first-order Trotterized product-ansatz strategy as in the N$_2$ calculation, with the Hartree--Fock determinant as the initial state and L-BFGS parameter optimization.

Compared with N$_2$, the tUCCSD-VQE energies for C$_2$ show a noticeably larger and more systematic deviation from the DMRG reference, even near the equilibrium geometry. As shown in the error panel, the deviation is already approximately 10--20 kcal/mol in the equilibrium region and increases again at stretched C--C distances, approaching 30 kcal/mol. This behavior reflects the substantially stronger multireference character of C$_2$, whose low-lying electronic states cannot be described quantitatively by a single-reference tUCCSD ansatz. In particular, the near-degeneracy of several valence configurations requires a more flexible treatment of electron correlation than that provided by single and double excitation generators defined relative to a single Hartree--Fock reference.

Nevertheless, the tUCCSD-VQE curve correctly reproduces the overall shape of the DMRG potential energy surface and remains significantly below the Hartree--Fock curve over the entire range of C--C bond lengths. This indicates that the ansatz still captures a large fraction of the correlation energy, even though it is not sufficiently expressive to reach quantitative agreement with DMRG for this strongly correlated system. Therefore, the discrepancy should be interpreted primarily as an ansatz limitation rather than a limitation of the Pauli-operator backend. The non-monotonic error around the bond length of 1.5--1.7~\AA\ may arise from the ordering dependence of the first-order Trotterized UCCSD ansatz. Such ordering effects have been discussed previously in the literature,\cite{grimsley2020trotterized} but a systematic investigation lies beyond the scope of the present work.

It is worth noting that the latter benchmark is particularly demanding from a computational perspective. The C$_2$/cc-pVDZ active space corresponds to a symmetry-reduced FCI dimension of $1.8\times10^{10}$, as summarized in Table~\ref{tab:binsim_csf_timing}. The successful evaluation of the full dissociation curve therefore tests not only the representational capability of a chemically motivated tUCCSD ansatz, but also the efficiency of BinSim in repeatedly applying large Pauli operators and evaluating energy expectation values in a high-dimensional correlated-electron Hilbert space.

\subsection{ADAPT-VQE Simulation of Molecular Torsion}

A fixed UCCSD ansatz may contain redundant operators and may become less effective in providing an accurate and compact description of strong electron correlation as the electronic structure evolves along a reaction coordinate. We therefore further examine whether BinSim can support adaptive ansatz construction, where the variational circuit is grown iteratively by selecting the most relevant excitation operators from a predefined operator pool. As a representative test, we computed the torsional potential energy curve of ethylene, C$_2$H$_4$, in the STO-3G basis using adaptive derivative-assembled pseudo-Trotter (ADAPT) VQE with a GSD excitation operator pool.\cite{grimsley2019adapt} This example provides a more flexible benchmark than a fixed-ansatz calculation because both the operator selection and the repeated optimization of an increasingly deep product ansatz must be performed along the potential energy surface.

In this calculation, the reference state was initialized from the Hartree--Fock determinant, and the selected operators were exponentiated and applied sequentially in a product form. The iteration procedure was terminated when $(\langle \hat{H}^2\rangle-\langle \hat{H}\rangle^2)/ \langle \hat{H}\rangle^2  < 10^{-7}$. At each adaptive step, the continuous variational parameters were optimized using L-BFGS with a More--Thuente line search, with convergence thresholds of $g_{\rm tol}=10^{-6}$ and $f_{\rm tol}=10^{-10}$. The active space is (16e,14o). The FCI reference energies were computed with PySCF. The errors are reported as energy deviations relative to FCI in kcal/mol.

\begin{figure}[!htb]
\centering
\includegraphics[width=0.95\linewidth]{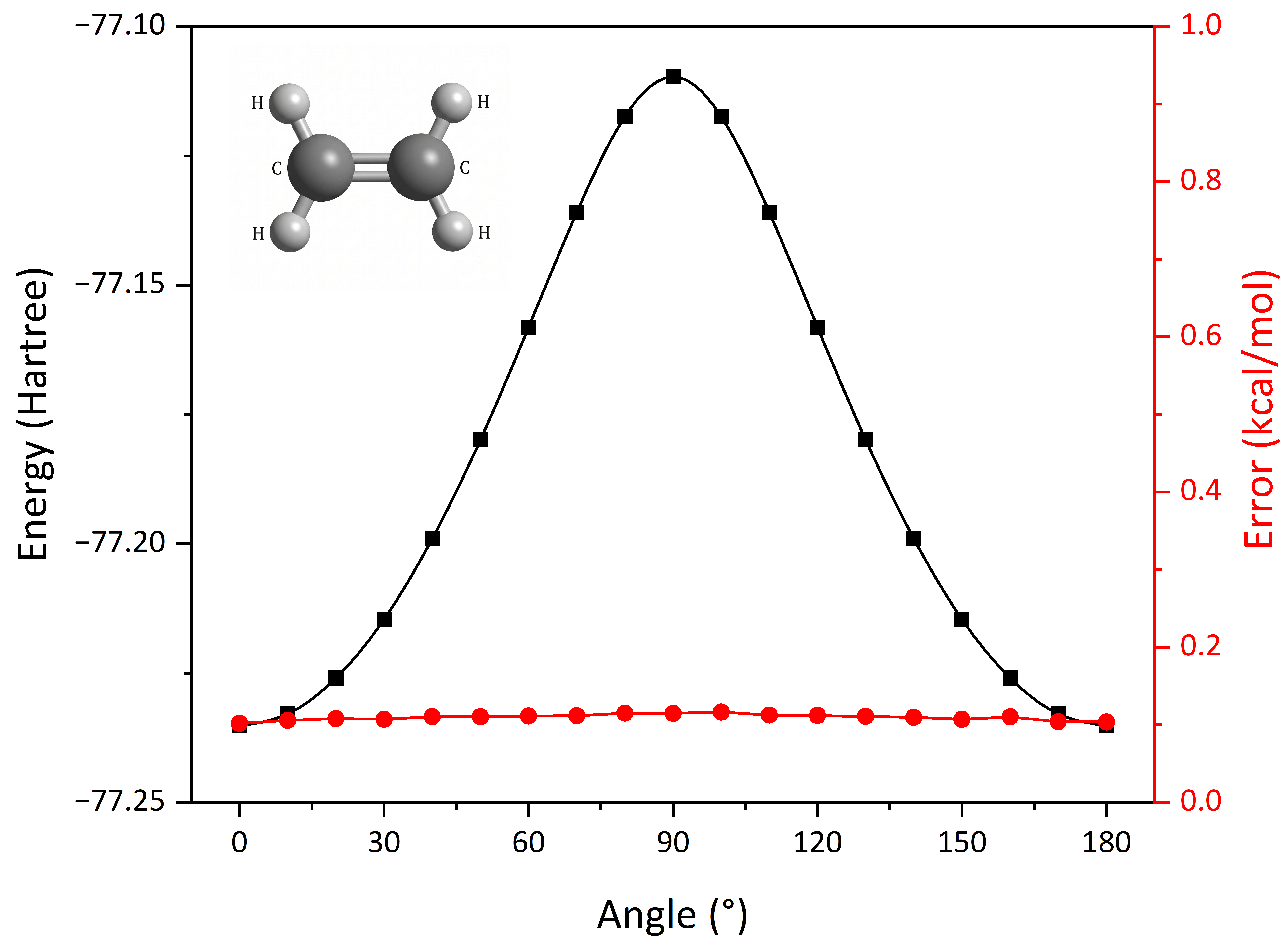}
\caption{Torsional potential energy curve of ethylene, C$_2$H$_4$, in the STO-3G basis. The black curve and black squares show the ADAPT-VQE energy as a function of the C--C torsional angle, plotted on the left energy axis. The red curve and red circles show the absolute energy error relative to FCI, plotted on the right axis in kcal/mol. }
\label{fig:ethylene_adapt_uccgsd}
\end{figure}

Figure~\ref{fig:ethylene_adapt_uccgsd} presents the ADAPT-VQE torsional potential energy curve of ethylene obtained with a GSD operator pool. The total energy increases from the planar geometry to a maximum near $90^\circ$, where the $\pi$ bond is strongly disrupted, and then decreases nearly symmetrically toward the equivalent planar structure at $180^\circ$. This profile is consistent with the expected torsional barrier associated with rotation around the C--C double bond. More importantly, the ADAPT-VQE energies remain uniformly close to the FCI reference throughout the scan, with absolute errors of approximately $0.1$ kcal/mol across all torsional angles. The error is well below the commonly used chemical-accuracy threshold of 1 kcal/mol, including the strongly distorted geometries near $90^\circ$.

The nearly flat error profile indicates that the adaptively constructed ansatz provides a balanced description along the entire reaction coordinate. This is particularly important for torsional motion, where the character of the electronic structure changes from a closed-shell-like planar configuration to a more strongly correlated twisted geometry. Unlike a fixed UCCSD ansatz, whose expressivity may vary substantially with geometry, ADAPT-VQE selects operators according to the instantaneous variational requirements of the wave function and then yields a geometry-consistent level of accuracy.

\subsection{Real-time Quantum Dynamics Simulations}

We also utilize BinSim as the backend for real-time quantum dynamics simulations. As a representative model, we considered an open-boundary TFIM,
\begin{equation}
     \hat{H}
     =
    J \sum_{i=1}^{N-1}
     \sigma_i^z \sigma_{i+1}^z
     +
    h \sum_{i=1}^{N}
     \sigma_i^x ,
 \end{equation}
with $J=1/4$, $h=1$ and $N=20$. The parameterized circuit employs the layered ansatz~\cite{Barison2021pVQD},
\begin{align}
   C(\mathbf{w})
   &=
   \prod_{\ell=1}^{L}
   c_\ell(\mathbf{w}_\ell), \\
    c_\ell(\mathbf{w}_\ell)
    &=
    \left[
    \prod_{i=1}^{N}
    R_{\alpha_\ell}^{(i)}(w_{i,\ell})
    \right]
    \left[
    \prod_{j=1}^{N-1}
    \exp\left(
    -i w_{j,\ell}
    \sigma_j^z\sigma_{j+1}^z
    \right)
    \right],
\end{align}
where
\begin{equation}
   R_{\alpha}^{(i)}(w)
    =
    \exp(-iw\sigma_i^\alpha),
    \qquad
    \alpha\in\{x,y\}.
\end{equation}
Here, $\mathbf{w}_\ell$ denotes the variational parameters in the $\ell$-th layer, and the number of layers $L$ corresponds to the circuit depth $d$. 
In the following calculations, each layer contains 20 single-qubit rotation parameters and 19 nearest-neighbor entangling parameters, giving 39 parameters per layer. Therefore, the ansatz depths $L=1,2,4,$ and $8$ correspond to 39, 78, 156, and 312 variational parameters, respectively. The time evolution was performed using the time-dependent variational principle (TDVP)\cite{McLachlan1964,yuan2019theory}, rather than the projected variational quantum dynamics (p-VQD) approach.~\cite{Barison2021pVQD}. The exact reference curves were generated by fourth-order Runge--Kutta (RK4) integration. The time step is 0.05.

\begin{figure}[!htb]
\centering
\includegraphics[width=1\linewidth]{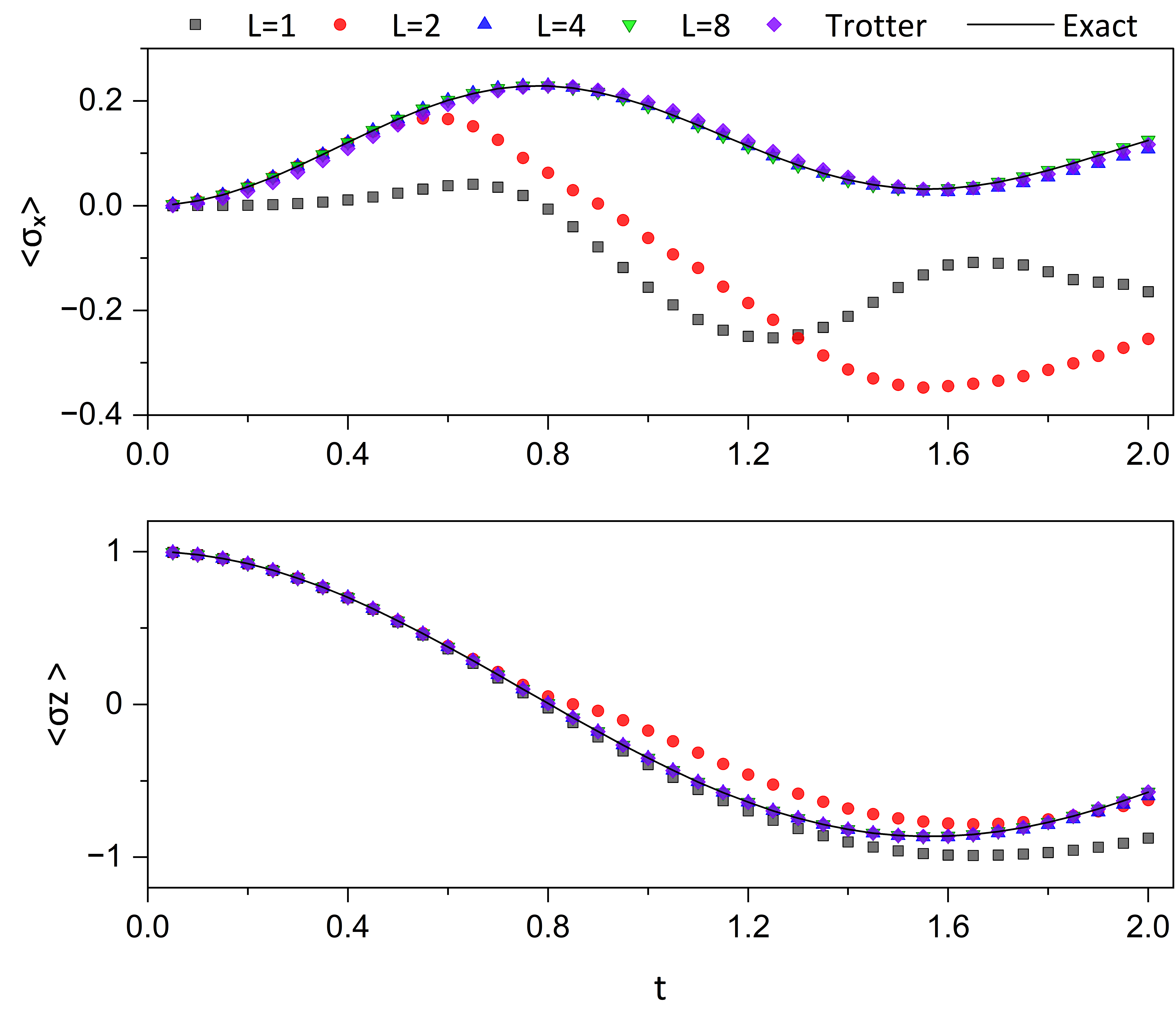}
\caption{Real-time quantum dynamics simulation of the open-boundary transverse-field Ising model. The upper panel shows the time evolution of the total transverse magnetization $\langle\sigma_x\rangle$, and the lower panel shows the longitudinal magnetization $\langle\sigma_z\rangle$. Colored markers are TDVP results obtained with different ansatz circuit depths, $L=1,2,4,$ and $8$. The black dashed curve is the exact reference obtained by RK4 propagation.}
\label{fig:ising_tdva}
\end{figure}

Figure~\ref{fig:ising_tdva} shows a clear and systematic improvement of the real-time dynamics with increasing circuit depth. The shallowest ansatz, $L=1$, captures only the short-time behavior and rapidly deviates from the exact RK4 reference, indicating that the corresponding variational manifold is insufficient to describe the subsequent entanglement growth and nontrivial spin rotations. Increasing the depth to $L=2$ extends the reliable simulation window, particularly for the longitudinal magnetization $\langle\sigma_z\rangle$, although noticeable deviations remain at later times and are more pronounced for the transverse magnetization $\langle\sigma_x\rangle$. In contrast, the $L=8$ results are nearly indistinguishable from the exact reference over the entire time interval, demonstrating convergence of the variational dynamics with respect to ansatz depth. First-order Trotterized time evolution also achieves high accuracy, although it is slightly less accurate than the $L=8$ variational results. Moreover, the circuit depth required by Trotterized time evolution increases linearly with the evolution time.

This depth dependence can be understood from the structure of the transverse-field Ising Hamiltonian. The interaction terms $\sigma_i^z\sigma_{i+1}^z$ and the transverse-field terms $\sigma_i^x$ do not commute, so real-time evolution simultaneously generates local spin rotations, many-body correlations, and phase coherence across the chain. A shallow circuit cannot faithfully represent this evolving structure once entanglement and phase correlations have built up. Adding more layers introduces additional rotational and entangling degrees of freedom, allowing the TDVP trajectory to remain closer to the exact dynamics. The different convergence behavior of $\langle\sigma_x\rangle$ and $\langle\sigma_z\rangle$ further indicates that transverse observables are more sensitive to the expressivity of the ansatz in this model. Overall, these benchmarks validate the TDVP implementation and demonstrate that the proposed Pauli-algebra backend can efficiently support the repeated operator applications required in real-time variational quantum dynamics.

\subsection{Computational Efficiency on Multicore CPU and GPU}

\begin{figure}[!htb]
\centering
\includegraphics[width=0.95\linewidth]{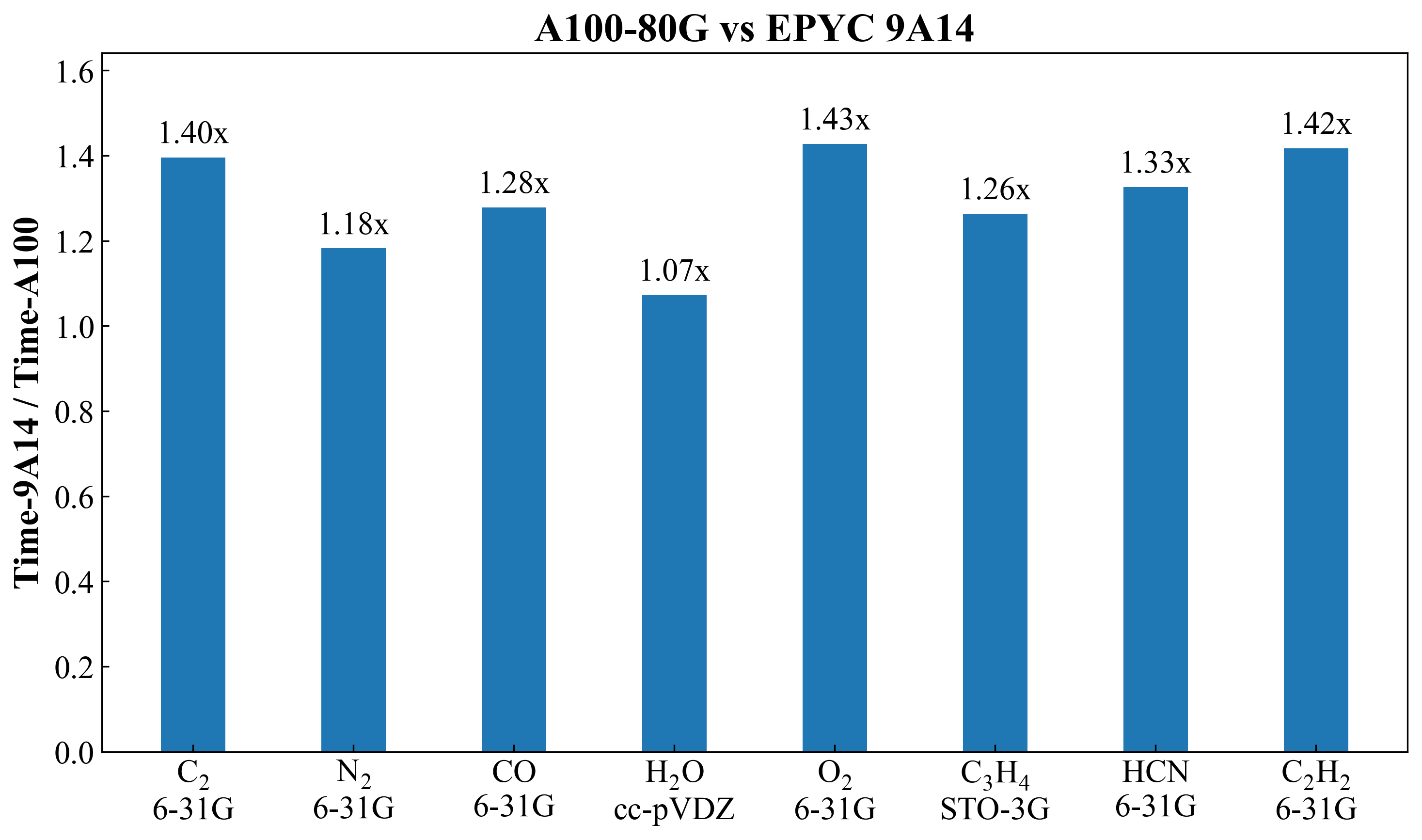}
\caption{Comparison of the wall-clock time for a single energy evaluation $E=\langle \Psi|\hat{H}|\Psi\rangle$ between one NVIDIA A100 80GB GPU and a dual-socket AMD EPYC 9A14 CPU node using 192 threads. The speedup is defined as Time-9A14 / Time-A100, which are the timings on the CPU and GPU, respectively.}
\label{fig:gpu_E_benchmark}
\end{figure}

To further evaluate the portability of BinSim across hardware architectures, we compared the wall-clock time of a single variational energy evaluation: $E=\langle \Psi|\hat{H}|\Psi\rangle$,
between a GPU implementation using one NVIDIA A100 80GB card and a CPU implementation using a dual-socket AMD EPYC 9A14 node with 192 threads. The speedup shown in Fig.~\ref{fig:gpu_E_benchmark} is defined as
${t_{\mathrm{9A14}}}/{t_{\mathrm{A100}}}$, so that values larger than 1 mean that the GPU is faster.

As shown in Fig.~\ref{fig:gpu_E_benchmark}, the GPU implementation outperforms the 192-thread CPU implementation for all tested systems.
These results show that the GPU version of BinSim can deliver a consistent performance advantage even when compared against a highly parallel CPU baseline using 192 threads. At the same time, the acceleration is moderate rather than dramatic, which suggests that the energy-evaluation kernel is not a purely dense floating-point workload. Instead, it still contains a significant amount of irregular data access, sparse operator application, and coefficient accumulation, all of which reduce the gap between GPU and CPU performance. In other words, the CPU implementation is already highly optimized, and the GPU benefit mainly comes from higher memory bandwidth and stronger throughput for batched bitwise and arithmetic operations.

The system dependence of the speedup also provides useful insight. For C$_2$/6-31G, O$_2$/6-31G and C$_2$H$_2$/6-31G, the GPU achieves the largest improvement, reaching 1.43$\times$, which indicates that these workloads provide enough arithmetic intensity and parallelism to better utilize the GPU hardware. In contrast, the nearly identical performance for H$_2$O/cc-pVDZ suggests that, for some systems, the cost may be dominated by memory movement, reduction overhead, or other operations that do not map as efficiently to the GPU architecture. Nevertheless, even in this least favorable case, the GPU still remains fully competitive with the 192-thread CPU execution.

Overall, Fig.~\ref{fig:gpu_E_benchmark} demonstrates that BinSim can efficiently exploit both many-core CPUs and modern GPUs. This hardware flexibility is important for practical large-scale variational simulation, since the energy evaluation $E=\langle \Psi|\hat{H}|\Psi\rangle$ is one of the most frequently repeated kernels in VQE and ADAPT-VQE workflows. The results indicate that a single A100 80GB GPU can already match or exceed the performance of a high-end 192-thread CPU node for this task.

\section{Conclusion}
\label{sec:conclusion}

We have developed a general computational framework for large-scale Pauli-algebra operations by combining binary symplectic encoding, Pauli-$(X,Z)$-mask-based acceleration, and memory-efficient parallelization. This framework supports high-performance operator--state multiplication that naturally cover dynamically expanding, fixed sparse, and effectively dense simulation regimes, providing a unified computational backend for Hamiltonian construction, iterative eigensolvers, variational optimization, and analytical-gradient evaluation on modern multicore CPUs and GPUs.

Beyond improving the efficiency of individual quantum simulation tasks, this work demonstrates that the design of the Pauli-algebra engine is itself a critical component of practical quantum computing for chemistry and materials science. Many near-term quantum algorithms spend a substantial fraction of their computational cost on classical preprocessing, operator manipulation, parameter optimization, and observable evaluation. By exploiting the exact algebraic structure of Pauli operators together with the sparsity and symmetry inherent in electronic-structure problems, the present framework substantially reduces these overheads without sacrificing generality. The resulting performance enables simulations in chemically relevant active spaces that would otherwise be difficult to access using conventional sparse-matrix representations.

An important feature of the framework is that it is algorithm-independent. The same operator infrastructure can be shared by VQE, quantum dynamics, Krylov-based eigensolvers, error-mitigation protocols, and other hybrid quantum--classical workflows, avoiding redundant implementations across different applications. As quantum hardware continues to improve, the computational bottleneck is expected to shift increasingly toward classical optimization and operator processing. Efficient classical infrastructures of this kind will therefore remain essential even in the fault-tolerant era, where increasingly large Hamiltonians and measurement workloads must still be managed efficiently.

Future developments will focus on transforming the present implementation from a high-performance Pauli-algebra engine into a broader algorithmic platform for quantum chemistry. One immediate direction is the systematic assessment of large-scale VQE and adaptive VQE algorithms, including their scaling behavior, optimization robustness, and limitations in strongly correlated active spaces. Another important direction is to extend the same operator infrastructure to excited states, periodic systems, and distributed heterogeneous computing environments, where the cost of Hamiltonian construction and repeated operator--state multiplication often becomes a central bottleneck. In this broader context, the present work provides a practical computational bridge between electronic-structure theory and quantum-algorithm design, facilitating the development, validation, and deployment of next-generation quantum chemistry methods on both classical simulators and quantum hardware.

\section{Appendix}

The comparisons in Table~\ref{tab:sim_comparison} used C$_2$H$_4$/STO-3G with an active space of (16e,14o). The UCCSD operator pool was reduced by D$_{2h}$ point-group symmetry, yielding 520 variational parameters; all timings were obtained with random circuit parameters. After the Jordan--Wigner transformation, the full Hilbert-space dimension is $2^{28}$. By contrast, ffsim and fqe operate in a sparse fermionic Fock space of dimension $\binom{14}{4}\binom{14}{4}\approx 1.2\times10^8$ and exploit D$_{2h}$ symmetry only to reduce the operator pool, not the determinant space; BinSim additionally reduces the CI dimension through point-group symmetry blocking.

For Qiskit-aer~2.5.0, the UCCSD ansatz was decomposed via \texttt{LieTrotter} (reps$\,{=}\,1$), flattened with \texttt{UCC.flatten}, and transpiled with optimization\_level$\,{=}\,3$, yielding a circuit depth of 99\,714. The simulation used the \texttt{AerSimulator} statevector backend. Because Qiskit-aer represents the Hamiltonian internally as a sparse CSC matrix, the Hv operation is not supported with this backend.

For ffsim~0.0.83, the Hv backend is \texttt{libfci.FCIcontract\_2es1} (OpenMP-enabled), while the Trotter backend is \texttt{ffsim.\_lib.contract\_fermion\_operator\_into\_buffer} (single-thread). For fqe~0.3.0, both the Hv (\texttt{Wavefunction.apply}) and Trotter (\texttt{FQEFactorizedUCC.apply}) backends support OpenMP.

\begin{acknowledgement}
This work is supported by Innovation Program for Quantum Science and Technology (2021ZD0303306), the National Natural Science Foundation of China (22422304, 22073086, 21825302, 22288201, 22393913 and 22303090), the Strategic Priority Research Program (XDB0450101) and the robotic AI-Scientist platform of the Chinese Academy of Sciences, Anhui Initiative in Quantum Information Technologies (AHY090400).
\end{acknowledgement}

\section*{Code availability}
The code used to perform the numerical simulations presented in this paper is publicly
available at https://github.com/halili138/BinSim.git

\section*{Notes}
The authors declare no competing financial interest.

\bibliography{qc}

\end{document}